\documentclass{article}
\usepackage[pdftex]{graphicx}
\usepackage{amsmath}
\usepackage{amssymb}
\usepackage{cite}

\usepackage{cite}

\begin{document}
\newcommand{\be}{\begin{equation}}
\newcommand{\ee}{\end{equation}}
\newcommand{\bm}{\boldmath}
\newcommand{\ds}{\displaystyle}
\newcommand{\bea}{\begin{eqnarray}}
\newcommand{\eea}{\end{eqnarray}}
\newcommand{\ba}{\begin{array}}
\newcommand{\ea}{\end{array}}
\newcommand{\arcsinh}{\mathop{\rm arcsinh}\nolimits}
\newcommand{\arctanh}{\mathop{\rm arctanh}\nolimits}
\newcommand{\bc}{\begin{center}}
\newcommand{\ec}{\end{center}}

\newcommand{\beqn}{\begin{eqnarray}}
\newcommand{\eeqn}{\end{eqnarray}}
\newcommand{\beqnn}{\begin{eqnarray*}}
\newcommand{\eeqnn}{\end{eqnarray*}}
\newcommand{\hr}{\hat\rho}
\def\a{\alpha}
\def\b{\beta}
\def\g{\gamma}
\def\l{\lambda}
\def\s{\sigma}
\def\vf{\varphi}
\def\ep{\epsilon}
\def\vep{\varepsilon}

\begin{center} {\Large \bf
%\begin{tabular}{c}
Excitation of the classical electromagnetic field in a cavity containing a thin slab with a time-dependent conductivity%\\[-1mm]
%\end{tabular}
 } \end{center}

\begin{center} {\bf
V. V. Dodonov and A. V. Dodonov
}\end{center}

%\medskip

\begin{center}
{\it
Instituto de F\'isica, Universidade de Bras\'ilia, Caixa Postal 04455, 70919-970  Bras\'ilia,  DF, Brasil
}
%\smallskip

E-mail:~~~vdodonov@fis.unb.br\\
\end{center}

\begin{abstract}\noindent

An exact infinite set of coupled ordinary differential equations, describing the evolution of the modes of the classical electromagnetic field 
inside an ideal cavity,
containing a thin slab with the time-dependent conductivity $\sigma(t)$ and dielectric permittivity $\vep(t)$, is derived for the dispersion-less media.
This problem is analyzed in connection with the attempts to simulate the so called Dynamical
Casimir Effect in three-dimensional electromagnetic cavities, containing a thin semiconductor slab, periodically illuminated
by strong laser pulses.
Therefore it is assumed that functions $\sigma(t)$ and $\delta\vep(t)=\vep(t)-\vep(0)$ are different from zero during short time
intervals (pulses) only. The main goal is to find the conditions, under which the initial nonzero classical field could be amplified
after a single pulse (or a series of pulses). 
Approximate solutions to the dynamical equations are obtained in the cases of ``small'' and ``big'' maximal values of
the functions $\sigma(t)$ and $\delta\vep(t)$.
 It is shown, that the single-mode approximation,
used in the previous studies, can be justified in the case of  ``small'' perturbations. But the initially excited
field mode cannot be amplified in this case, if the laser pulses generate free carriers inside the slab. 
The amplification could be possible, in principle, for extremely
high maximal values of conductivity and the concentration of free carries (the model of ``almost ideal conductor''), 
created inside the slab, under the crucial condition, that the function $\delta\vep(t)$ is {\em negative}. 
This result follows from a simple approximate analytical solution, and it is confirmed by exact numerical calculations.
However, the evaluations show, that the necessary energy of laser pulses must be, probably, unrealistically high. 
 \end{abstract}

%\medskip


\section{Introduction}

The aim of this paper is to study the evolution of the classical electromagnetic field inside an ideal cavity,
containing a thin slab with the time-dependent conductivity and dielectric permittivity. This problem seems to be interesting by itself,
as soon as it was not investigated in detail earlier (as far as we know), but our main motivation stems from its connection with
the experiments on the Dynamical Casimir Effect (DCE) in cavities. Originally, this effect was thought as
a possibility of creating the electromagnetic field quanta from the vacuum initial state in macroscopic systems
(e.g., cavities) with moving boundaries. Comprehensive reviews on this subject can be found, e,g., in
\cite{D-rev1,DD-rev2,VD-PS10,DAR-11,PNA-12}. 
In the original formulation the effect was predicted by Moore \cite{Moore}, 
although the name DCE was coined much later \cite{Yabl,Sas}.

It was known from the very beginning, that the effect should be extremely small for non-relativistic motions of
boundaries. For example, in the simplest model of the cavity DCE, one can suppose that a single field mode is excited due to the change of its
eigenfrequency, caused by the displacement of the boundary. Neglecting the intermode interaction, 
one can use the model of a single harmonic oscillator with time-dependent frequency. Suppose that some field
quadrature oscillates as $x(t) =\cos(\omega t)$ for $t<0$, but the frequency changes abruptly to $\Omega$
at $t=0$, returning to the initial value $\omega$ at $t=\tau$. (This model simulates an instantaneous jump of
boundary at $t=0$ and its coming back to the initial position at $t=\tau$; see, e.g., \cite{Sas}). Then  $x(t)=\cos(\Omega t)$ 
for $0<t<\tau$ (due to the continuity of function $x(t)$ and its derivative), 
whereas for $t>\tau$ we have $x(t) =A\cos(\omega t) +B\sin(\omega t)$, with
\[
A = \cos(\omega \tau)\cos(\Omega \tau) +\frac{\Omega}{\omega}\sin(\omega \tau)\sin(\Omega \tau),
\qquad
B =  \sin(\omega \tau)\cos(\Omega \tau) -\frac{\Omega}{\omega}\cos(\omega \tau)\sin(\Omega \tau).
\]
The ratio of energies $\dot{x}^2 +\omega^2 x^2$ at the instants $t>\tau$ and $t<0$ is equal to
\be
W= 1 +\sin^2(\Omega\tau) \frac{\Omega^2 -\omega^2}{\omega^2}.
\ee
We see that the energy increases, if $\Omega >\omega$ (i.e., for contracting cavities). 
This increase is very small under the realistic conditions, $|\Omega-\omega|/\omega \ll 1$,
but it can be amplified, in principle, by means of periodical repetitions of the process with a proper choice of the
repetition period, using the parametric resonance effect \cite{DK92,DK96,Lamb,Dal99,Plun00}.
 
Unfortunately, it is extremely difficult to realize the necessary 
 resonance conditions for periodic motions of real cavity walls at the current experimental level
\cite{VD-PS10,DK96}. Therefore different ideas to {\em simulate\/} the real motion of boundaries by fast variations of their
material properties were proposed independently by several authors \cite{Yabl,Man91,Loz}.
They were realized, partially, in  recent experiments \cite{Wilson,Para}, where cavities were replaced by superconducting quantum
circuits, and time-dependent boundary conditions, simulating periodic boundary displacements, were achieved due to
fast variations of parameters of the Josephson contacts (SQUIDs). (Note that the initial idea to use the Josephson contacts for
the simulation of DCE in quantum circuits was put forward as far back as in \cite{Man91}.)

However, the dream to observe the ``true'' DCE in {\em cavities\/} has not been realized until now.
It was suggested some time ago \cite{Padua,Padua05} to simulate the motion of ideal boundaries by changing the conductivity
of a thin semiconductor slab near the metallic wall, using a chain of short laser pulses.
Indeed, it was verified in the preliminary experiment \cite{Padua}, that the illuminated semiconductor slab behaves as
a metal, from the point of view of the high reflectivity. Therefore it was supposed that a periodic creation of a highly conducting
film near the slab surface, followed by the recombination of carriers, could be equivalent to periodic displacements of the metallic boundary,
with the displacement amplitude of the order of the slab thickness (up to a few millimeters), which is much bigger than the maximal possible
amplitude of displacements of real vibrating surfaces at the frequencies belonging to the GHz range (a few nanometers \cite{DK96}).
The experiment was named MIR (Motion/Mirror Induced Radiation).
However, despite that many difficulties were overcome \cite{Pad-Cas60,Brag09,Brag-RSI11,Brag-reentr}, the effect was not observed.

One of severe obstacles in the MIR scheme was clear from the very beginning \cite{DD-rev2}. Theoretical predictions
\cite{DK92,DK96,Lamb,Dal99,Plun00} were made under the important assumption that cavity walls were {\em ideal conductors},
i.e., neglecting energy losses in  the cavity. The peculiarity of semiconductor materials is that losses can be small 
at the initial stage of the excitation process (when the conductivity is very small) and at the middle stage of high (almost metallic)
conductivity, but they can be high enough during the inevitable intermediate stages of the excitation/recombination process,
when the conductivity is not very small, but not very big, too. Consequently, losses are inevitable. 
This observation stimulated the development of theoretical models of quantum damped oscillator with
time-dependent parameters, which could describe the evolution of quantized field modes in ``bad'' cavities
with time-dependent damping coefficients
\cite{DD-rev2,VD-PS10,DD05,D05,DD06a,DD06b,DD06J,Dod-PRA2009,Dod-RMF}.

In addition, it was known, that even in ideal cavities,
 the intermode interaction could significantly diminish the rate of photon generation  \cite{DK96,Croc1,AVD-01,DD-PLA12}.
Therefore the models constructed in \cite{DD-rev2,VD-PS10,DD05,D05,DD06a,DD06b,DD06J,Dod-PRA2009,Dod-RMF} 
were based on the assumption, that the
intermode interaction could be neglected, even in the case of strong changes of conductivity, so that the problem could be reduced
to that of a single quantum damped nonstationary oscillator. In this single-mode regime, it was predicted that an exponential
growth of the mean photon number from the initial vacuum or thermal states could be possible under certain tough 
restrictions on the duration, periodicity, and energy of laser pulses, depending on the mobility of carriers and recombination
times in the semiconductor. 
Actually, the estimations performed in the frameworks of the models 
\cite{DD-rev2,VD-PS10,DD05,D05,DD06a,DD06b,DD06J,Dod-PRA2009,Dod-RMF} 
should be considered as  {\em upper limits\/} for a possible number of photons, generated due to the DCE in cavities with
semiconductor slabs.
However, although some predictions of the models were confirmed (with a surprising accuracy) in the series of last experiments 
of the MIR project, the amplification of the initial probe signal in the cavity was not observed: losses always prevailed the 
parametric amplification effect \cite{MIR-rep}.

 Therefore, it seems interesting to understand, whether the failure was due to the incorrect choice of material and
other parameters, or there exist more fundamental reasons (missed in the previous models), that preclude a possibility of observing the DCE in cavities
with semiconductor materials? A related (but more technical) question is, whether a single-mode description of the problem 
could be justified, or such an approach is a non-adequate oversimplification?
To answer these questions, we consider the evolution of {\em classical\/} electromagnetic field inside a 
cylindrical (rectangular) cavity
with fixed ideal walls, containing a slab with time-dependent conductivity $\sigma$ and dielectric permittivity $\vep$,
which can depend on the longitudinal (along the cylinder axis) coordinate only.
The choice of such a simple geometry enables us to calculate all necessary coefficients of equations explicitly and exactly.
The aim is to see, whether the energy of {\em classical field\/} can increase or not, when the conductivity of the slab
goes rapidly to some high value and then returns to the initial zero value.
Considering the classical electrodynamics, we can avoid a complicated problem of the field quantization in time-dependent dissipative
media, since all calculations can be performed on the basis of Maxwell's equations. On the other hand, if the DCE is possible, 
then the number of quanta should grow exponentially, and the initial quantum field inevitably would become classical.
Consequently, if the DCE is possible, then the initial classical signal must be amplified.

\section{Basic equations}

The dynamical Maxwell equations in the non-conductive media read (we use the Gauss system of units and
 consider non-magnetic materials, so that ${\bf B} \equiv {\bf H}$)
\be
\mbox{rot}{\bf B} =   \frac{1}{c}\frac{\partial {\bf D}}{\partial t}, 
\label{Max0B}
\ee
\be
\mbox{rot}{\bf E} = -\,\frac{1}{c}\frac{\partial {\bf B}}{\partial t}.
\label{Max0E}
\ee
If the medium is inhomogeneous, but isotropic and non-dispersive, then ${\bf D}({\bf r},t)=\vep({\bf r}){\bf E}({\bf r},t)$,
with a real function $\vep({\bf r}) \ge 1$. In such a case, there exists the complete set of discrete eigenmodes, possessing the monochromatic time
dependence in the form
\be
{\bf E}({\bf r},t) = {\bf E}_{{\bf n}}({\bf r}) \exp\left(-i\omega_{{\bf n}}t\right), \qquad
{\bf B}({\bf r},t) = -i{\bf B}_{{\bf n}}({\bf r}) \exp\left(-i\omega_{{\bf n}}t\right).
\ee
The real time independent functions ${\bf E}_{{\bf n}}({\bf r})$ and ${\bf B}_{{\bf n}}({\bf r})$ satisfy the equations
\be
\mbox{rot}{\bf B}_{{\bf n}} =   \vep({\bf r}) \omega_{{\bf n}}{\bf E}_{{\bf n}}({\bf r})/c,
\label{Max0omB}
\ee
\be
\mbox{rot}{\bf E}_{{\bf n}} = \omega_{{\bf n}}{\bf B}_{{\bf n}}({\bf r})/c.
\label{Max0omE}
\ee
Actually, the ``vector'' index ${\bf n}$ is a set of three integers $(l,m,n)$.

An immediate consequence of Eqs. (\ref{Max0omB}) and (\ref{Max0omE}) is the orthogonality of mode functions with different eigenvalues
\cite{Vainshtein}:
\be
\int \vep({\bf r}) {\bf E}_{\bf n}({\bf r}){\bf E}_{\bf m}({\bf r})dV= 
\int {\bf B}_{\bf n}({\bf r}){\bf B}_{\bf m}({\bf r})dV =0, \qquad {\bf n} \neq {\bf m},
\label{ortog}
\ee
and the equality
\be
\int \vep({\bf r}) {\bf E}_{\bf n}^2({\bf r})dV= \int {\bf B}_{\bf n}^2({\bf r})dV.
\ee
The integration in the equations above is performed over the total volume of the cavity. It is assumed that the cavity walls are ideal conductors.

It seems reasonable to suppose  that the set of functions ${\bf E}_{{\bf n}}({\bf r})$  and ${\bf B}_{{\bf n}}({\bf r})$ is not only orthogonal, but
also {\em complete\/} for the given fixed geometry and boundary conditions (although we did not find the proof of this statement in
available textbooks, except for the case of homogeneous dielectric media inside the cavity). Under this assumption
(which can be crucial: see the discussion in Sec. \ref{sec-discuss}), arbitrary electromagnetic field  vectors
can be written as
\be
{\bf E}({\bf r}, t) = \sum_{{\bf n}}{\bf E}_{{\bf n}}({\bf r}) f_{{\bf n}}( t),
\qquad {\bf B}({\bf r}, t) = \sum_{\bf n}{\bf B}_{{\bf n}}({\bf r}) g_{{\bf n}}( t).
\label{EBexp}
\ee
Using the normalization of the eigenmode field vectors in the form
\be
\int dV \vep({\bf r})\left[{\bf E}_{{\bf m}}({\bf r})\right]^2 = 
 \int dV \left[{\bf B}_{{\bf m}}({\bf r})\right]^2 = 
8\pi,
\label{normcond}
\ee
we can write the total energy of the field as
\be
W = \frac1{8\pi}\int dV\left({\vep\bf E}^2 + {\bf B}^2\right)
= 
\sum_{{\bf n}}\left( f_{\bf n}^2 + {g}_{\bf n}^2 \right).
\label{toten}
\ee

%it is easy to see from Eq. (\ref{Max1}) that $g_{{\bf n}}( t)=\dot{f}_{{\bf n}}( t)$.

%Moreover, $\int {\bf D}_{\bf n}{\bf B}_{\bf m}dV =0$ for any sets of numbers ${\bf n}$ and ${\bf m}$.  ???

Now let us consider the cavity containing a medium with the time-dependent dielectric permittivity $\vep({\bf r}) +\delta\vep({\bf r},t)$ 
and conductivity $\sigma({\bf r},t)$. This assumption, implying the neglect of dispersion, seems reasonable for microwave fields 
with frequencies below THz.
Then Eq. (\ref{Max0B}) should be replaced by
\be
\mbox{rot}{\bf B} =  \frac{4\pi \sigma}{c} {\bf E} +  \frac{1}{c}\frac{\partial }{\partial t}\left[(\vep + \delta\vep){\bf E}\right]. 
\label{Max1}
\ee
Using Eqs. (\ref{EBexp}) and (\ref{Max0omB}),  we obtain the equation
\be
\sum_{\bf n}  {\bf E}_{\bf n}\left[ \vep\left(\dot{f}_{\bf n} - g_{\bf n}\omega_{\bf n}\right)  + \delta\vep\dot{f}_{\bf n}
 + f_{\bf n} \left(4\pi\sigma   + \partial \delta\vep/{\partial t}\right) \right] =0.
\ee
Now we can integrate the scalar product of this equation with vector ${\bf E}_{\bf m}$ over the cavity volume.
Taking into account (\ref{ortog}) and (\ref{normcond}),  we arrive at the infinite set of coupled ordinary differential equations
\be
\dot{f}_{\bf m} = \omega_{\bf m}g_{\bf m} -\sum_{\bf n} \alpha_{\bf mn}f_{\bf n} - \frac{d}{dt}
\sum_{\bf n} \beta_{\bf mn}f_{\bf n},
\label{dotf}
\ee
where
\be
\alpha_{\bf mn}(t) =  \frac12\int dV \sigma({\bf r},t){\bf E}_{\bf m}({\bf r}){\bf E}_{\bf n}({\bf r}), \quad
\beta_{\bf mn}(t) = \int dV \frac{\delta\vep({\bf r},t)}{8\pi}{\bf E}_{\bf m}({\bf r}){\bf E}_{\bf n}({\bf r}).
\label{albe}
\ee
The same scheme gives rise to the following consequence of Eq. (\ref{Max0E}):
\be
\dot{g}_{\bf m} = -\omega_{\bf m} f_{\bf m}.
\label{dotgm}
\ee
Excluding $g_{\bf m}$, we arrive at the system of coupled second order ordinary differential equations
\be
\ddot{f}_{\bf m} + \omega_{\bf m}^2 f_{\bf m} = -\frac{d}{dt}\sum_{\bf n} \alpha_{\bf mn}f_{\bf n} - \frac{d^2}{dt^2}
\sum_{\bf n} \beta_{\bf mn}f_{\bf n}.
\label{ddotf}
\ee

\section{Small variations of $\sigma$ and $\vep$}
\label{sec-small}

If coefficients $\alpha_{\bf mn}$ and $\beta_{\bf mn}$ are small, then one can apply the perturbation
theory, replacing the functions $f_{\bf n}(t)$ in the right-hand side of Eq. (\ref{ddotf}) by their unperturbed (initial, for $t<0$) values
$A_{\bf n}\cos\left(\omega_{\bf n} t +\phi_{\bf n}\right)$, i.e., considering the right-hand side of Eq. (\ref{ddotf}) as some known ``force'' $f(t)$. 
We suppose that functions $\alpha_{\bf mn}(t)$ and $\beta_{\bf mn}(t)$ equal zero for
$t<0$.  Then the known solution of the classical forced oscillator
\be
F(t)= A\cos\left(\omega t +\phi\right) + \omega^{-1}\int_0^{t} d\tau
\sin\left[\omega (t-\tau)\right] f(\tau)
\ee
results in the following approximate solutions to Eq.  (\ref{ddotf}) at $t>t_*$, where $t_*$ is the time instant when $\sigma$ and
$\delta\vep$ return to the initial zero values (we used integrations by parts):
\be
f_{\bf m}(t) = A_{\bf m}\cos\left(\omega_{\bf m} t +\phi_{\bf m}\right)  - u_{\bf m}\cos\left(\omega_{\bf m} t\right) 
+ v_{\bf m}\sin\left(\omega_{\bf m} t\right),
\ee
\be
u_{\bf m}=
\sum_{\bf n} A_{\bf n} \int_0^{t_*} d\tau
\cos\left(\omega_{\bf n} \tau +\phi_{\bf n}\right)\left[\omega_{\bf m} \beta_{\bf mn}(\tau)\sin\left(\omega_{\bf m}\tau\right)
+\alpha_{\bf mn}(\tau)\cos\left(\omega_{\bf m} \tau\right)\right],
\label{um}
\ee
\be
v_{\bf m}=
\sum_{\bf n} A_{\bf n} \int_0^{t_*} d\tau
\cos\left(\omega_{\bf n} \tau +\phi_{\bf n}\right)\left[\omega_{\bf m} \beta_{\bf mn}(\tau)\cos\left(\omega_{\bf m}\tau\right)
-\alpha_{\bf mn}(\tau)\sin\left(\omega_{\bf m} \tau\right)\right].
\label{vm}
\ee

 We have $g_{\bf m}= \dot{f}_{\bf m}/\omega_{\bf m}$ at $t<0$ and $t>t_*$. Consequently,
the energy $W_{\bf m}$ of the ${\bf m}$th mode at $t\ge t_*$, defined according to Eq. (\ref{toten}), equals 
\be
W_{\bf m} = A_{\bf m}^2 -2A_{\bf m} \left(u_{\bf m} \cos\phi_{\bf m} +v_{\bf m} \sin\phi_{\bf m}\right) +u_{\bf m}^2 +v_{\bf m}^2.
\ee
Using approximate formulas (\ref{um}) and (\ref{vm}), we have to neglect the second order terms $u_{\bf m}^2 +v_{\bf m}^2$.
In this approximation, the energy difference $\Delta W_{\bf m} = W_{\bf m} - A_{\bf m}^2$ equals
\be
\Delta W_{\bf m} = -2A_{\bf m} \sum_{\bf n} A_{\bf n} \int_0^{t_*} d\tau
\cos\left(\omega_{\bf n} \tau +\phi_{\bf n}\right)\left[\omega_{\bf m} \beta_{\bf mn}(\tau)\sin\left(\omega_{\bf m}\tau +\phi_{\bf m}\right)
+\alpha_{\bf mn}(\tau)\cos\left(\omega_{\bf m} \tau +\phi_{\bf m}\right)\right].
\ee

Suppose that one of coefficients, $A_{\bf M}$, is much bigger than all other coefficients $A_{\bf n}$ with ${\bf n} \neq {\bf M}$.
Then the energy change in the ${\bf M}$th mode is much bigger than in other modes, unless some specific initial conditions
(related to the initial phase $\phi_{\bf M}$) are satisfied:
\be
\Delta W_{\bf k} = -2A_{\bf k}  A_{\bf M} \int_0^{t_*} d\tau
\cos\left(\omega_{\bf M} \tau +\phi_{\bf M}\right)\left[\omega_{\bf k} \beta_{\bf kM}(\tau)\sin\left(\omega_{\bf k}\tau +\phi_{\bf k}\right)
+\alpha_{\bf kM}(\tau)\cos\left(\omega_{\bf k} \tau +\phi_{\bf k}\right)\right],
\ee
\be
\Delta W_{\bf M} = -W_{\bf M} \int_0^{t_*} d\tau
\left[\omega_{\bf M} \beta_{\bf MM}(\tau)\sin\left(2\omega_{\bf M}\tau +2\phi_{\bf M}\right)
+2\alpha_{\bf MM}(\tau)\cos^2\left(\omega_{\bf M} \tau +\phi_{\bf M}\right)\right].
\label{DelWM}
\ee

Note that the diagonal coefficients $\alpha_{\bf mm}$  are non-negative for arbitrary non-negative functions
$\sigma(z,t) $, due to Eq. (\ref{albe}). 
Consequently, the nonzero small conductivity always gives a negative
contribution to the energy change of the leading ${\bf M}$th mode, in agreement with our intuition. The energy change due to
the change of dielectric constant can be positive or negative, depending on the time $t_*$ and phase $\phi_{\bf M}$. These 
parameters can be
adjusted to give a maximal positive change of energy, proportional to the initial energy of the selected mode. This observation indicates a
possibility of exponential increase of energy of the selected mode due to the parametric resonance.

\section{Parametric resonance with a monochromatic excitation}
\label{sec-paramres}

Let us suppose that coefficients $\alpha_{\bf MM}(t) \equiv \alpha(t)$ and $\beta_{\bf MM}(t) \equiv \beta(t)$ 
perform harmonic oscillations at the frequency
close to $2\omega_{\bf MM} \equiv 2\omega$. Since $\sigma \ge 0$, we consider the model with $\alpha(t) =\alpha_0 \sin^2(\Omega t)$,
where $\Omega \approx \omega$ (so that $\alpha(0) =0$).  
Having in mind applications to the cases of photo-excited semiconductor media,
we may suppose that $\beta(t)=\beta_0 \sin^2(\Omega t)$, where constant coefficients $\alpha_0$ and $\beta_0$ are
proportional  to the maximal concentration of additional carriers, created by the periodic laser illumination. The sign of 
coefficient $\beta_0$ can be either positive or negative, whereas $\alpha_0 \ge 0$. Under this assumption, it is reasonable
to assume, that all other coefficients $\alpha_{\bf mn}(t)$ and $\beta_{\bf mn}(t)$ with ${\bf m} \neq {\bf n}$ have the same
time dependence, differing in the amplitude coefficients. If all coefficients in the right-hand side of Eqs. (\ref{ddotf})
are small, we may look for the solution in the form of quasi-harmonic oscillations with {\em slowly varying amplitudes}:
\be
f_{\bf m}(t)= A_{\bf m}(t) \exp\left(i\omega_{\bf m} t\right) + B_{\bf m}(t) \exp\left(-i\omega_{\bf m} t\right)
\label{fAB}
\ee
Following \cite{Land}, we put these expressions in Eqs. (\ref{ddotf}), neglecting the second order derivatives $\ddot{A}_{\bf m}$
and $\ddot{B}_{\bf m}$, believing that they have higher orders of smallness. After that, we multiply the equations obtained
by the factors $ \exp\left(\pm i\omega_{\bf m} t\right) $ and perform averaging over fast oscillations. Assuming that the
spectrum of eigenfrequencies is not equidistant, or more precisely, that the differences 
$\omega_{\bf m} \pm \omega_{\bf n} \pm 2\Omega$ are not close to zero for all values ${\bf m} \neq {\bf n}$,
one can see that the infinite set of coupled equations (\ref{ddotf}) can be reduced to two coupled equations for the
amplitudes $A_{\bf M}(t)=A(t)$ and $B_{\bf M}(t)=B(t)$:
\be
\dot{A} = \frac18\left(\alpha_0 + i\omega\beta_0\right)\left(B e^{2i\delta t} -2A\right), \qquad
\dot{B} = \frac18\left(\alpha_0 - i\omega\beta_0\right)\left(A e^{-2i\delta t} -2B\right).
\label{dotAB}
\ee
Here $\delta =\Omega -\omega$ is considered as a small quantity, so that the terms proportional to $\alpha_0 \delta$
and $\beta_0 \delta$ are neglected, as well as 
 small terms proportional to the products of $\alpha_0$ and $\beta_0$ by $\dot{A}$ and $\dot{B}$.
The substitutions $A(t)=\tilde{A}(t)\exp(i\delta t)$ and $B(t)=\tilde{B}(t)\exp(-i\delta t)$ transform (\ref{dotAB}) to
the system of equations with constant coefficients
\be
\dot{\tilde{A}} = \frac18\left(\alpha_0 + i\omega\beta_0\right)\tilde{B}- \frac14\left(\alpha_0 + i\omega\beta_0 + 4i\delta\right)\tilde{A}, \quad
\dot{\tilde{B}} = \frac18\left(\alpha_0 - i\omega\beta_0\right)\tilde{A} + \frac14\left( i\omega\beta_0 + 4i\delta -\alpha_0 \right)\tilde{B}.
\label{dottilAB}
\ee
Looking for solutions proportional to $\exp(\lambda t)$, we obtain the values
\be
\lambda_{\pm} = -\frac{\alpha_0}{4} \pm \frac18\sqrt{\alpha_0^2 + \omega^2 \beta_0^2 
-4\left(\omega\beta_0 +4\delta \right)^2}.
\ee
The condition of parametric amplification is $\lambda_{+} >0$, i.e.,
\be
\omega^2 \beta_0^2 > 4\left(\omega\beta_0 +4\delta \right)^2 + 3\alpha_0^2.
\label{cond-amp}
\ee
Consequently, the best choice of the frequency detuning is $\delta= -\omega\beta_0/4$. This quantity is different from zero,
because the average value of function $\beta(t)$ over the period of oscillations equals $\beta_0/2 \neq 0$.

\section{Cylindrical cavity with a homogeneous dielectric slab }

To evaluate possible values of parameters $\alpha_0$ and $\beta_0$ in a photo-excited semiconductor slab,
we consider an arbitrary cylindrical cavity with the longitudinal dimension  $L_z\equiv L$, so that $0<z<L$.
Let us suppose that a dielectric slab with a {\em constant\/} value of $\vep >1$ occupies the space $0<z<L_s <L$, whereas
the rest part of the cavity is empty ($\vep=1$).
Then the consequence of Eqs. (\ref{Max0B}) and (\ref{Max0E}) is the Helmholtz equation for any component of the
electric field:
\be
\Delta {\bf E}_{\bf n} + {\bf k}_{\bf n}^2\tilde{\vep}{\bf E}_{\bf n} =0,
\qquad
\tilde{\vep} = \left\{
\begin{array}{ll}
\vep, & 0<z<L_s
\\
1, & L_s < z <L
\end{array}
\right., \qquad  {\bf k}_{\bf n}^2 \equiv \left(\omega_{\bf n}/c\right)^2.
\ee
This equation admits solutions in the factorized form with respect to the longitudinal coordinate $z$ and transverse vector
${\bf r}_{\perp} \equiv (x,y)$.
  In order to be close to the geometry of re-entrant cavities, used in the MIR experiments, we consider the TM modes. This means that
	$B_z \equiv 0$, while the $E_z$ component is nonzero. Then the components of vector ${\bf E}({\bf r})$ 
can be written as follows (we omit the mode index ${\bf n}$ in all cases when this cannot result in a confusion), 
\be
E_z = N{\bf k}_{\perp}^2\Phi({\bf r}_{\perp})\psi(z),  \quad
E_x = N\frac{\partial\Phi}{\partial x}\frac{d\psi}{dz}, \quad
E_y = N\frac{\partial\Phi}{\partial y}\frac{d\psi}{dz},
\label{Phi}
\ee
where $N$ is the normalization factor. Functions $\Phi({\bf r}_{\perp})$ and $\psi(z)$ satisfy the equations
\be
\Delta_{\perp}\Phi + {\bf k}_{\perp}^2 \Phi =0,
\ee
\be
\psi^{\prime\prime} +\kappa^2(z) \psi =0, \qquad \kappa^2(z) =
\left\{ \begin{array}{ll}
\kappa_s^2 \equiv {\bf k}_n^2 \vep -k_{\perp}^2, & 0<z < L_s
\\
k_n^2 \equiv {\bf k}_n^2 -k_{\perp}^2, & L_s < z < L
\end{array}
\right.
\label{eqpsi}
\ee
with suitable boundary conditions. One can check, in particular, that
the condition $\mbox{div}{\bf E}=0$ is satisfied automatically inside and outside the slab.

We suppose that a thin conducting film of the effective thickness $\delta_s \ll L_s$, containing periodically photo-excited carriers, 
is formed near the surface
$z=L_s$. If the functions $\sigma({\bf r},t)$ and $\delta\vep({\bf r},t)$ do not depend on the transversal coordinates $x$ and $y$,
then one can deduce from Eq. (\ref{albe}) the relation
\be
\alpha(t)/\beta(t) = 4\pi \int \sigma(z,t)dz \left[\int \delta\vep(z,t) dz\right]^{-1}.
\label{al-be}
\ee
In the simplest case of free photo-excited carriers, described by the Drude formula, we have the known expressions for the conductivity and
variation of dielectric function at the frequency $\omega$:
\be
\sigma({\bf r},t)= \frac{n({\bf r},t) e^2 \gamma}{m_{ef}\left(\omega^2 + \gamma^2\right)}, \qquad
\delta\vep({\bf r},t)= -\,\frac{4\pi n({\bf r},t) e^2 }{m_{ef}\left(\omega^2 + \gamma^2\right)},
\label{Drude}
\ee
where $n({\bf r},t)$ is the concentration of free carriers, $e$ the electron charge, $m_{ef}$ the effective mass, and $\gamma$ is the collision frequency.
Comparing (\ref{al-be}) and (\ref{Drude}), we obtain the ratio
\be
\left|\alpha_0/\beta_0\right| = \gamma.
\ee
But the non-dispersive highly conducting regime exists under the condition $\gamma \gg \omega$. This means that the condition of parametric 
resonance (\ref{cond-amp}) cannot be fulfilled for small variations of  $\sigma$ and $\vep$ in the non-dispersive regime.

\section {Very thin highly conducting slab}

Now let us consider the case, when the conductivity inside a thin slab rapidly increases to some very big value and then again rapidly
decreases to zero. 
If functions $\sigma({\bf r},t)$ and $\delta\vep({\bf r},t)$ do not depend on the transverse coordinates $x$ and $y$,  then
coefficients $\alpha_{\bf mn}$ and $\beta_{\bf mn}$ are diagonal with respect to the first indices $l,m$ of the vector labels 
${\bf m}=(l,m,n)$ and ${\bf n}=(l',m',n')$. This is obvious for the rectangular cavity with $|x| <L_x/2$ and $|y|<L_y/2$, 
where function $\Phi\left({\bf r}_{\perp}\right)$,
defined in Eq. (\ref{Phi}), has the form 
\be
\Phi(x,y)= \cos\left(k_l x\right) \cos\left(k_m y\right), \qquad
k_l =(1+2l)\pi/L_x, \quad k_m =(1+2m)\pi/L_y.
\label{Phirec}
\ee 
This means that modes with different first two indices $l$ and $m$ are not coupled in the case involved.
Therefore we consider hereafter the excitation of field modes with fixed indices $l=m=0$,
introducing the simplified notation $f_m \equiv f_{00m}$, $g_m \equiv g_{00m}$, $\omega_m \equiv \omega_{00m}$.
Then Eqs.  (\ref{dotf}) and (\ref{dotgm}) can be written in the following vector form:
\be
\dot{\bf f} = {\Omega}{\bf g} -{\cal A}{\bf f} -\frac{d}{dt}({\cal B} {\bf f})  ,
\label{dotvecf}
\ee
\be
\dot{\bf g} = -{\Omega}{\bf f},
\label{dotvecg}
\ee
where ${\bf f} \equiv \left(f_{0}, f_{1}, f_{2}, \ldots\right)$, ${\bf g} \equiv \left(g_{0}, g_{1}, g_{2}, \ldots\right)$, and
$\Omega$ is the diagonal matrix: $\Omega = \mbox{diag}\left(\omega_m\right)$.
 Infinite-dimensional symmetrical matrices ${\cal A}(t)$ and ${\cal B}(t)$ have the following elements:
\be
a_{mn}(t) = \tilde{N}_m \tilde{N}_n\int_0^{L_s}dz \sigma(z,t)\psi_m(z)\psi_n(z), \qquad
b_{mn}(t) = \frac{\tilde{N}_m \tilde{N}_n}{4\pi}\int_0^{L_s}dz \delta\vep(z,t)\psi_m(z)\psi_n(z),
\label{amnbmn}
\ee
where $ \tilde{N}_m$ and $ \tilde{N}_n$ are constant factors, determined by the normalization of the mode eigenfunctions.

An immediate consequence of Eqs. (\ref{dotvecf}) and (\ref{dotvecg}) is the equation for the time derivative of the total energy 
of all the modes $W= {\bf f}^2 +{\bf g}^2$:
\be
\frac12 \frac{dW}{dt} = - {\bf f}{\cal A}{\bf f} - {\bf f}\frac{d}{dt}({\cal B} {\bf f}).
\label{dWdt}
\ee
The frequency matrix $\Omega$ does not enter the right-hand side of this equation, since this matrix is {\em symmetric}.
Since $\sigma(z,t) \ge 0$, then one can easily check, using the definition (\ref{amnbmn}), that matrix ${\cal A}$ is
{\em non-negatively definite}, i.e., ${\bf f}{\cal A}{\bf f} \ge 0$ for any vector ${\bf f}$.
This means that $dW/dt \le 0$, if ${\cal B} =0$. Consequently, the variations of the dielectic permittivity are crucial
for the possibility of the field amplification: without such variations, the total energy will always decrease, due to the losses
caused by the nonzero conductivity.

Eq. (\ref{dotvecf}) can be re-written as 
\be
\dot{\bf f} = \left[I +{\cal B}(t)\right]^{-1}\left[{\Omega}{\bf g} -\left({\cal A} +\dot{\cal B}\right){\bf f}\right]  ,
\label{dotvecfinv}
\ee
where $I$ is the identity matrix. Here we meet some technical difficulty, caused by the necessity to calculate the inverse matrix
$\left[I +{\cal B}(t)\right]^{-1}$, whose dimensionality can be rather high.

It is remarkable, that this difficulty can be overcome, if
we make the assumption (which is quite reasonable in view of the MIR experiments performed in Padova), 
that functions $\sigma(z,t)$ and $\delta\vep(z,t)$ are concentrated in a very thin film
(of thickness much smaller than $L_s$) near the surface $z=L_s$ of the dielectric slab. 
In such a case, matrices ${\cal A}(t)$ and ${\cal B}(t)$ can be represented as
\be
{\cal A}(t) =a(t)\Psi, \quad {\cal B}(t) =b(t)\Psi, 
\label{AB}
\ee
where
\be
a(t) = \int_0^{L_s} \sigma(z,t) dz/L, \quad b(t) = \int_0^{L_s} \delta\vep(z,t) dz/L,
\label{atbt}
\ee
while the elements of symmetric matrix $\Psi$ are
\be 
\Psi_{mn} = L \tilde{N}_m \tilde{N}_n \psi_m(L_s)\psi_n(L_s) \equiv \Psi_m \Psi_n.
\label{defPsin}
\ee 
Explicit expressions for the coefficients $\Psi_n$ in the case of rectangular cavity are given in the Appendix.
An immediate important consequence of Eq. (\ref{AB})  is the commutativity of matrices ${\cal A}(t)$ and ${\cal B}(t')$
(and their time derivatives) for any values of time $t$ and $t'$. Using this commutativity and
making the substitution 
\be
{\bf f} = \left[I +{\cal B}(t)\right]^{-1}{\bf y}, \qquad {\bf f}(0) = {\bf y}(0), \quad  {\bf f}(t_*) = {\bf y}(t_*),
\ee
one can reduce Eq. (\ref{dotvecfinv})  to the form
\be
\dot{\bf y} = {\Omega}{\bf g} -\left[I +{\cal B}(t)\right]^{-1}{\cal A} {\bf y}.
\label{doty}
\ee
A remarkable property of matrix $\Psi$ is the factorization of its elements (therefore $\det\Psi=0$). Its immediate important consequence
is the set of identities
\be
\Psi^2 = \rho \Psi, \qquad \Psi^n = \rho^{n-1} \Psi,
\label{Psi2}
\ee
\be
\rho = \mbox{Tr}(\Psi) =\sum_{k=0}^{\infty} \Psi_k^2 .
\label{rho}
\ee
The convergence of series (\ref{rho}) is shown in the Appendix. %\ref{ap-conv}. 
Assuming that matrix ${\cal B}$ is ``small'', we can write a formal expansion
\be
\left[I +{\cal B}\right]^{-1} = I +\sum_{n=1}^{\infty}(-b)^n \Psi^n =
 I -b\Psi \sum _{n=0}^{\infty}(-b\rho)^n
= I -\frac{b}{1+b\rho}\Psi.
\label{expan}
\ee
But it is easy to check that the equality of the first and last expressions in (\ref{expan}) is, in fact, an algebraic identity, following from Eq. (\ref{Psi2}).
Consequently,
\[
\left[I +{\cal B}\right]^{-1}{\cal A}= a\Psi  -\frac{ba}{1+b\rho}\Psi^2 =  \frac{a}{1+b\rho}\Psi.
\]
Therefore Eq. (\ref{doty}) can be written as
\be
\dot{\bf y} = -\,\frac{a(t)}{1+b(t)\rho}\Psi {\bf y} + {\Omega}{\bf g},
\label{doty1}
\ee
whereas Eq. (\ref{dotvecg}) takes the form
\be
\dot{\bf g} = -\Omega \left[I -\frac{b(t)}{1+b(t)\rho}\Psi\right] {\bf y}.
\label{dotg1}
\ee

\section{Heuristic approximate analytical solution}

Since the matrices $\Psi$ and $\Omega$ do not commute (this can be easily verified), a reliable solution to the set of equations
(\ref{doty1}) and (\ref{dotg1}) can be found only numerically. This will be done in Sec. \ref{sec-num}.
But before doing this,  let us try to ``guess'' some qualitative properties of the solution, based on some kind of heuristic approach.
For this purpose, we note that 
 it is possible to obtain a simple approximate analytical solution, neglecting the term ${\Omega}{\bf g}$ in the
right-hand side of Eq. (\ref{doty1}). Of course, this trick cannot be justified in the generic case, but some weak justification can be
given for the special initial condition ${\bf g}={\bf 0}$, if function $a(t)$ can attain very big values (i.e., for almost perfectly
conducting slab in the time interval $0<t<t_*$. So, let us make this approximation and see, what can happen.

In this case, Eq. (\ref{doty1}) can be solved immediately:
\be
{\bf y}(t) = \exp\left[-\tilde\kappa(t)\Psi\right] {\bf y}(0),
\qquad \tilde\kappa(t) =\int_0^{t} \frac{ a(\tau) d\tau}{1+ \rho b(\tau)}.
\label{yty0}
\ee
The matrix exponential in (\ref{yty0}) can be easily calculated, taking into account Eq. (\ref{Psi2}):
\be
\exp(-\tilde\kappa \Psi) = \sum _{n=0}^{\infty}(-\tilde\kappa \Psi)^n/n! 
= I + \Psi\rho^{-1}\sum _{n=1}^{\infty}(-\tilde\kappa\rho)^n/n!
=  I - \Psi \left[1- e^{-\tilde\kappa\rho}\right]/\rho.
\ee
Consequently,
\be
y_m(t) = y_m(0) - \Lambda(t) \Psi_m Y , 
\label{fmt0}
\ee
where
\be
 Y = \sum_{n=0}^{\infty} \Psi_n f_n(0), \qquad 
\Lambda(t) = \left\{1 -\exp[-\kappa(t)]\right\}/\rho,
\qquad
\kappa(t) =\int_0^{t} \frac{\rho a(\tau) d\tau}{1+ \rho b(\tau)}.
\label{def-kap}
\ee

Let us suppose that, initially, the only mode with $m= M$ was excited. Then $Y= \Psi_M f_M(0)$, so that
\be
f_M(t_*) = f_M(0) \left[1-\Lambda_* \Psi_M^2\right] , 
\qquad   f_k(t_*) = -\Psi_k \Psi_M  \Lambda_*  f_M(0),
 \quad k \neq M,
\label{fM}
\ee
%\be
%\label{fMk}
%\ee
where
\be
  \Lambda_* \equiv  \Lambda(t_*) = \left\{1 -\exp[-\kappa_*]\right\}/\rho, \qquad \kappa_* \equiv \kappa(t_*).
\label{fmt*}
\ee
We see that the crucial parameter is the sign of coefficient $\Lambda_*$, which, in turn, coincides with the sign of coefficient $\kappa_*$.
If $\kappa_*>0$, then the amplitude of the initially excited mode diminishes. But if $\kappa_* <0$, then the field mode amplitude
can be amplified.

Using Eq. (\ref{fmt0}), we can obtain a simple formula for the change of the total electric energy of all modes 
for an arbitrary initial vector ${\bf f}(0)$
(the magnetic field energy should be neglected within the accuracy of the approximation ${\bf g}(0)={\bf 0}$ used in this section):
\be
W_{el}(t_*) - W_{el}(0) \equiv \sum_{m=0}^{\infty}\left[f_m^2(t_*) -f_m^2(0)\right]
= \Lambda_* Y^2(\Lambda_*\rho -2)  
= -Y^2 \left[1 -\exp\left(-2\kappa_*\right)\right]/\rho.
\label{Wneg}
\ee
We see that the sign of the total energy change depends on the sign of coefficient $\kappa_*$. If $\delta\vep>0$,
then $b(t)>0$ and $\kappa(t) >0$, so that no amplification can be expected in this case.
Note, however, that $\delta\vep$ is negative, if this quantity
can be evaluated with the aid of the Drude model (\ref{Drude}). Under this assumption, $b(t)<0$, as well. If $|b(\tau)|\rho < 1$
in the integrand of Eq. (\ref{def-kap}) (i.e., if the concentration of carriers, generated by the laser pulse, is not high enough), 
then $\kappa(t) >0$, and the total field energy decreases, obviously due to the losses, caused by the finite conductivity.
But the situation can be different, if the function  $|b(\tau)|\rho $ can exceed the unit value.
Therefore the heuristic model of this section indicates, that the necessary condition for the amplification can be the negativity
of $\delta\vep$ and the big absolute value of this function.
Actually, the case of big absolute values of the negative function $\rho b(\tau)$ needs some care, 
because the denominator $1+b(\tau)\rho$ in the integrand of the formula for $\kappa(t)$
 passes through zero value. But this seems not a big problem, because the solution found in this section is only an approximation. 
In the next section we show, that numerical solutions show a similar qualitative behavior of the field energy.

\section{Numerical solutions}
\label{sec-num}

In the numerical tests, we put formally $\omega_0=1$, $t_*=0.01$, $L_s/L=0.01$, and $\mu=1/2$ 
(this value corresponds to the cubical cavity: see Appendix), so that $\rho= 4.255$.  % $\rho= 4.255238272185344$.
We solved Eqs. (\ref{doty1}) and (\ref{dotg1}), using the Fortran program and  taking into account the first 600 modes, 
checking that  the variation of this number to 700 modes changed the results not more than by 0.3\% in the worst case
(for $E$ and the largest positive values of $\rho c_0$: see the notation below).
Being interested in the case of $\delta\vep<0$, we used the function $c(t)=-b(t)$. Remembering that this function must turn into zero
at $t=0$ and $t=t_*$, we considered a simple trial function $c(t)=c_0\sin(\pi t/t_*)$.
We supposed that only the first mode ($m=1$) was excited initially (since the frequency $\omega_0$ does not depend on the cavity
length $L$, the excitation of this mode cannot simulate the DCE), so that $f_n(0)=g_n(0)=0$ for $n \neq 1$.
The initial conditions for the first mode were chosen in the form $f_1(0)=\cos(\phi)$, $g_1(0)=\sin(\phi)$, so that both, the initial energy
of the first mode, and the total field energy, were equal to unity.
Below we present the plots of the partial, $E_1$, and total, $E$, energies at $t=t_*$, where
\be
E_n=f_n^2(t_*) +g_n^2(t_*), \qquad E=\sum_{n=0}^{\infty} E_n.
\ee

\begin{figure}[htb]
\vspace{-1cm}
%\begin{center}
\includegraphics[width=0.99\textwidth]{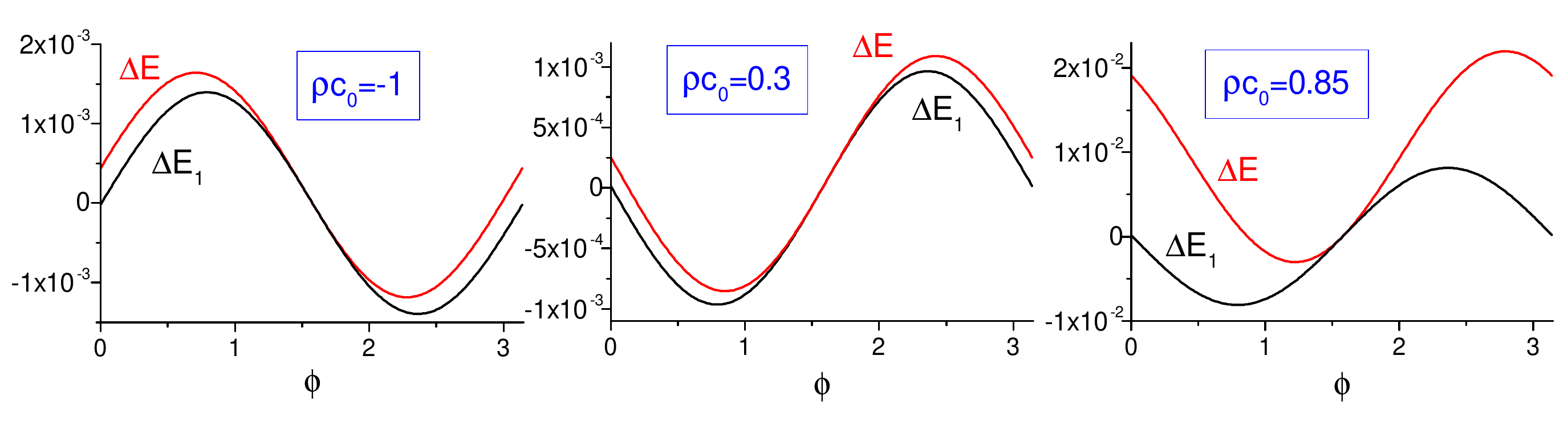} 
%\hspace{-3cm}%
%\hfill% 
%\includegraphics[width=0.5\textwidth]{test99.png} 
\vspace{-6mm}
\caption{\label{fig-gam0} $\Delta E_1$ and $\Delta E$ as functions of phase $\phi$ for $\gamma=0$. %$\rho c_0=0.1$ (left) and $\rho c_0=0.99$ (right).
}
\end{figure} 
Figure \ref{fig-gam0} shows the dependence of $\Delta E_1 =E_1-1$ and $\Delta E=E-1$ on the phase $\phi$ 
in the case of $\gamma=0$ (an ideal dielectric
without losses), for positive and negative values of the product  $\rho c_0$.
%=-0.99$ (this means that $\delta\vep >0$), $\rho c_0=0.2$, and $\rho c_0=0.99$ (i.e.,  $\delta\vep <0$).
We see that the possibility of the field amplification is strongly phase dependent, in a qualitative agreement with predictions of Secs.
\ref{sec-small} and \ref{sec-paramres}. We see also, that the single mode approximation works quite well for small perturbations
($\rho c_0=0.3$), and even for strong {\em positive\/} perturbations of the dielectric permittivity ($\rho c_0=-1$),
but it can be not very reliable for strong {\em negative\/} perturbations ($\rho c_0=0.85$), 
when $\Delta E_1$ and $\Delta E$ can take opposite signs.

\begin{figure}[hbt]
\vspace{-0.5cm}
\begin{center}
\includegraphics[width=0.7\textwidth]{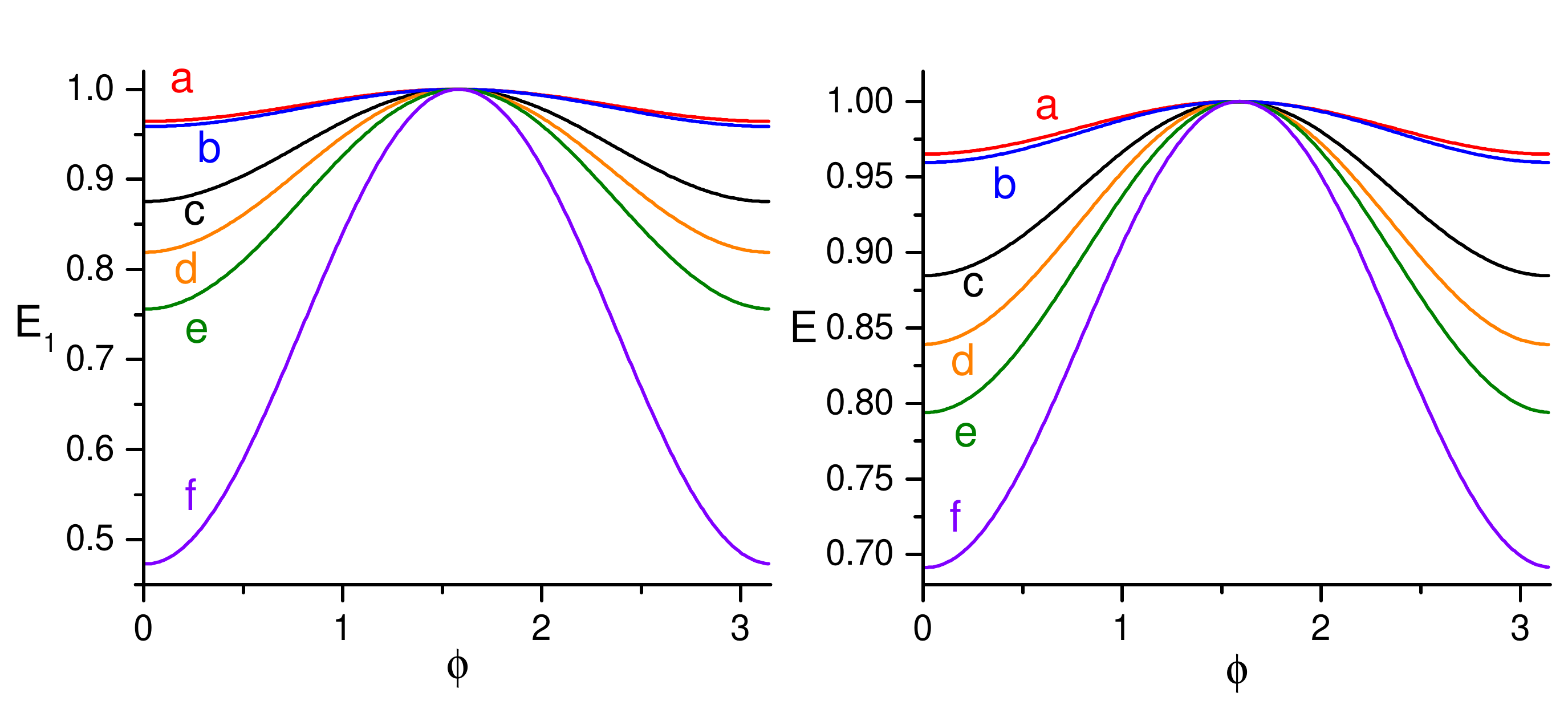} 
%\hspace{-3cm}%
%\hfill% 
%\includegraphics[width=0.5\textwidth]{test99.png} 
\vspace{-6mm}
\caption{\label{fig-gam100-} $E_1$ and $E$ as functions of phase $\phi$ for $\gamma=100$ and different values of $\rho c_0$.
The concrete values of this product are as follows, a: -0.1; b: 0.1; c: -0.5, d: -0.99; e: 0.5; f: 0.99.
}
\end{center}
\end{figure} 
Figure \ref{fig-gam100-} shows $E_1$ and $E$ as functions of phase $\phi$ for $\gamma=100$ and different values of $\rho c_0<1$.
No amplification is observed in this case, in accordance with predictions of the preceding section.

\begin{figure}[htb]
\begin{center}
\includegraphics[width=0.7\textwidth]{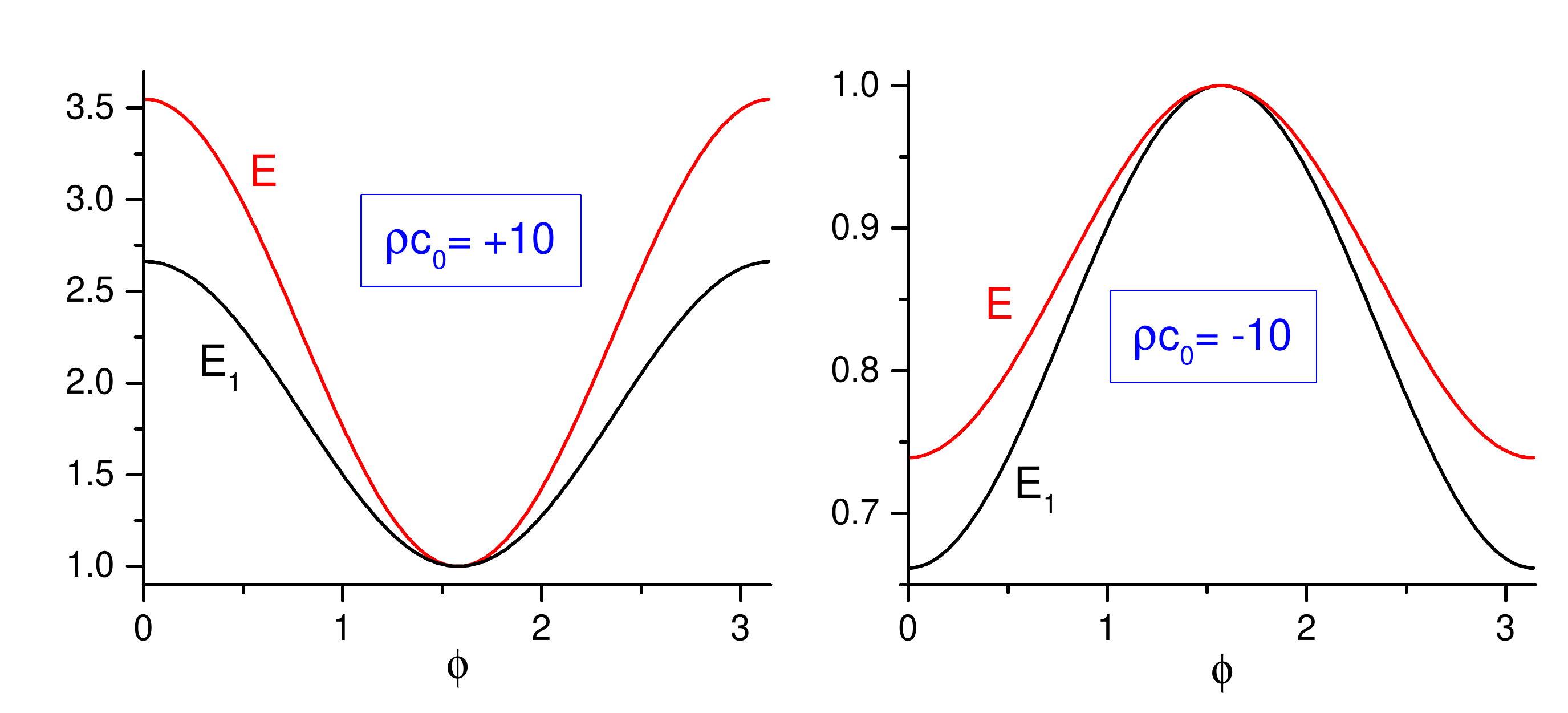} 
%\hspace{-3cm}%
%\hfill% 
%\includegraphics[width=0.5\textwidth]{test99.png} 
%\vspace{-4mm}
\caption{\label{fig-gam100+} $E_1$ and $E$ as functions of phase $\phi$ for $\gamma=100$, $\rho c_0=10$  (left), and 
$\rho c_0 = -10$ (right).
}
\end{center}
\end{figure}

The case of $\rho c_0=10$ is shown in Fig. \ref{fig-gam100+}.
To avoid the problem with zero value of denominator $1-\rho c(t)$ in Eqs. (\ref{doty1}) and (\ref{dotg1}), we suppose that
function $c(t)$ increases very rapidly to the value $c_0$ during a very short interval $0<t<\delta  \ll t_*$, then remains constant
and goes down to zero during a very short interval $t_*-\delta <t< t_*$. Integrating Eq. (\ref{dotvecf}) over time from $0$ to $\delta $,
we can neglect the integral $\int_0^{\delta } \left[{\Omega}{\bf g}(\tau) -{\cal A}(\tau){\bf f}(\tau)\right] d\tau$, assuming that $\delta  \to 0$,
whereas the functions in the integrand remain limited. Thus we get
${\bf f}(\delta) -{\bf f}(0) = {\cal B}(0){\bf f}(0) -{\cal B}(\delta){\bf f}(\delta)$.
Remembering that ${\cal B}(0)=0$ and ${\cal B}(\delta ) = -c_0 \Psi$, 
we obtain the new initial condition
\be
{\bf f}(\delta) =\left[I +{\cal B}(\delta)\right]^{-1}{\bf f}(0) = \left[I +\frac{c_0}{1 -c_0\rho}\Psi\right]{\bf f}(0).
\label{fdelta}
\ee
Assuming that $\dot{\cal B}=0$ for $\delta <t < t_* - \delta$, we find the vector ${\bf f}(t_*)$, solving the equation
\be
\dot{\bf f} = \left[I +\frac{c_0}{1 -c_0\rho}\Psi\right]\left[{\Omega}{\bf g} -{\cal A}{\bf f}\right]
\ee
with the initial condition (\ref{fdelta}). The coordinate form of this equation reads
\be
\dot{f}_m = \omega_m g_m  +\frac{c_0 \Psi_m}{1-\rho c_0} \sum_{n=0}^{\infty}\Psi_n \omega_n g_n 
 - \frac{\gamma |c_0| \Psi_m}{1-\rho c_0}  \sum_{n=0}^{\infty}\Psi_n f_n  .
\label{dotfmc0}
\ee
The second equation remains 
\be
\dot{g}_m = -\omega_m f_m.
\label{dotgmagain}
\ee
Finally, the value of ${\bf f}(t_*)$ is obtained by the integration of (\ref{dotvecf}) from $t_*-\delta$ to $t_* $:
\be
{\bf f}(t_*) =\left[I+{\cal B}(t_*-\delta)\right]{\bf f}(t_*-\delta) =
\left[I -c_0 \Psi\right]{\bf f}(t_*-\delta).
\label{ffin}
\ee
We took into account that ${\cal B}(t_*)=0$.
In the numerical calculations we used the limit $\delta \to 0$, integrating Eqs. (\ref{dotfmc0}) and (\ref{dotgmagain}) from $t=0$ to $t=t_*$,
using the initial and final conditions  (\ref{fdelta}) and (\ref{ffin}). In this scheme, it is assumed that ${\bf g}(\delta)={\bf g}(0)$ and 
${\bf g}(t_*-\delta)={\bf g}(t_*)$. 

All the plots show that $E_1=E=1$ for $\phi=\pi/2$. Actually, this is an accidental equality, due to the small chosen value of the pulse duration
$t_*=0.01$. We have checked, that for bigger values of $t_*$, the quantity $E_1$ is different from unity (although the difference is small).

%\section{Single-mode approximation}
%\label{sec-single}

\section{Discussion}
\label{sec-discuss}

Let us emphasize the main results obtained in this paper. We have derived exact equations governing the evolution of the
classical electromagnetic field inside the cylindrical (rectangular) cavity with ideal boundaries, when the conductivity $\sigma$ and
dielectric permittivity $\vep$ inside a thin slab, attached to the base of the cylinder, vary with time. If the perturbation of the field is small
(due to the smallness of $\sigma$ and $\delta\vep$), then the single-mode model can be justified. However, no amplification can be
achieved in this case for microwave fields (provided the temporal dispersion can be neglected), 
if $\sigma$ and $\delta\vep$ are due to the creation of {\em free carriers\/} inside the slab. This result can be considered as an
expected one.

The new result (which seems to be quite new) is the possibility to achieve (at least, in principle) the amplification effect
in the case of very big concentrations of the laser-created free carriers, 
under the strong condition, that not only the conductivity must be high, but the
time dependent dielectric permittivity must be also big in the absolute value, being {\em strongly negative}.
In other words, the semiconductor slab must become almost an ideal conductor with $\sigma \to \infty$ and  $\vep \to -\infty$.
The second condition can be thought as impossible, at the first glance, but it does not contradict the general physical principles.
Indeed, it was shown in \cite{Kirzh}, that the restriction on the admissible values of the static (or low-frequency) 
dielectric permittivity is not $\vep \ge 1$,
but $1/\vep \le 1$, which does not exclude possible negative values of $\vep$. Whether this condition can be fulfilled in
real semiconductors, is an open question (this is possible in the frameworks of the Drude model, but the validity of this model can be 
questioned in the case of high concentration of carriers). In any case, the results obtained show that the field amplification
certainly cannot be achieved, if $\delta\vep$ is positive or merely slightly negative.

Here we meet a subtle point, namely, the problem of passing through the zero value of the denominator $1+\rho b(t)$ in Eqs.
(\ref{doty1}) and (\ref{dotg1}) 
[or the determinant of matrix $I+{\cal B}(t)$ in the general case (\ref{dotvecfinv})]. We tried to avoid it, but probably this difficulty indicates
the incompleteness of the expansion (\ref{EBexp}) of the total electromagnetic field over the standing modes ${\bf E}_{\bf n}$ and
${\bf B}_{\bf n}$ in the case, when the function $\vep({\bf r}) +\delta\vep({\bf r},t)$ becomes negative. In this case, the surface
plasmon polariton waves can exist \cite{Rama05}. The corresponding EM field has the evanescent behavior, which is not
taken into account in the expansion (\ref{EBexp}).

Closing eyes on these subtleties, let us try to evaluate possible numerical values of parameters, that could give rise to the 
amplification of the non-evanescent part of the field.
Using Eqs. (\ref{Drude}) and (\ref{atbt}), we can estimate $b_{max}$ as
\be
|b_{max}| \sim \frac{4\pi E_p e^2}{E_g m_e\gamma^2 V_{cav} } ,
\label{bmax}
\ee
where $E_p$ is the laser pulse energy (which is assumed to be totally absorbed to produce the free carriers with the $100$\% efficiency),
$E_g$ is the energy gap of the semiconductor, and $V_{cav}$ is the cavity volume.
Let us consider, for the purpose of an evaluation, the cubical cavity with $\omega_0 = 10^{10}\,$s$^{-1}$. Then $L \sim 14\,$cm
and $\mu=1/2$, according to Eqs. (\ref{omegmu}) and (\ref{Nn}). Since we use the condition $\gamma \gg \omega_m$, let us take
$\gamma= 10^{12}\,$s$^{-1}$. Taking $E_g \sim 1.5\,$eV and $E_p\sim 100\,\mu$J \cite{Pad-Cas60}, we obtain
$|b_{max}| \sim 5\times 10^{-4}$. Since $\rho \approx 4$ for $\mu=1/2$, we have 
$|b_{max}| \rho \sim 2\times 10^{-3}$. Certainly, the field amplification is impossible for these parameters.
One has to increase the value of $|b_{max}| $ at least by three orders of magnitude.
However, the pulses of energy $0.1\,$J seem unrealistically strong, and they can simply burn the semiconductor sample.
Taking a higher frequency, say, $\omega_0= 10^{11}\,$s$^{-1}$, one can reduce the cavity volume $10^3$ times. However,
one should increase $\gamma$ in this case to $\gamma= 10^{13}\,$s$^{-1}$, so that the minimal energy of each pulse will be 
$0.01\,$J, which is still a big value.

Note that the effective volume of the gap in the reentrant cavity, used in the MIR experiment, is just about $1\,$cm$^3$.
However, formula (\ref{bmax}) cannot be applied to this geometry. One can expect, in principle, that the field amplification could be also
achieved in this case, for big enough energies of laser pulses, but theoretical predictions for the re-entrant cavity seem to be practically impossible,
because one has  to solve in this case Eqs. (\ref{dotf})--(\ref{dotgm}) for the functions $f_{\bf m}$ and $g_{\bf m}$, which
depend on three indices. The coefficients $\alpha_{\bf mn}$ and $\beta_{\bf mn}$ depend on $6$ indices in the general case, and 
one has to know all the mode functions with the sufficient accuracy to find these coefficients. It is quite probable, that the necessary minimal
pulse energy could occur unreasonably high and unachievable in real experiments.

\section*{Acknowledgments}

The partial support of the Brazilian funding agency CNPq is acknowledged. We thank Dr.  I. V. Voronich for some
numerical tests, which were useful at the initial stage of this study.
%\appendix
\renewcommand{\theequation}{A.\arabic{equation}}
\setcounter{equation}{0}

\section*{Appendix: Explicit expressions for some coefficients and the proof of convergence of series (\ref{rho})}
\label{ap-conv}

It is easy to check the convergence of series (\ref{rho}) for an empty cavity with $\vep=1$ everywhere. 
The presence of a thin slab of thickness $L_s\ll L$ with $\vep \sim 1$ does not change the field distribution significantly,
so that the approximation of the empty cavity seems reasonable for the evaluations below.
Taking $l=m=0$, we have the following explicit expressions for the electric field of the  mode ${\bf n}=(0,0,n)$:
\be
E_{z,{ n}}= N_n \cos\left(\pi x/L_x\right)\cos\left(\pi y/L_y\right)\cos\left(\pi n z/L\right), 
\label{solcos1}
\ee
\be
E_{x,{ n}}= N_n\left(n\mu L/L_x\right) \sin\left(\pi x/L_x\right)\cos\left(\pi y/L_y\right)\sin\left(\pi n z/L\right),  
\label{solcosx1}
\ee
\be
E_{y,{ n}}= N_n\left(n\mu L/L_y\right)  \cos\left(\pi x/L_x\right)\sin\left(\pi y/L_y\right)\sin\left(\pi n z/L\right),
\label{solcosy1}
\ee
\be
N_n^{-2} = \frac18 L_x L_y L\left(1+\mu n^2\right) \left(1+\delta_{n0}\right),  
\qquad \mu = \frac{L_x^2 L_y^2}{L^2\left(L_x^2 +L_y^2\right)}.
\label{Nn}
\ee
Coefficient $\mu$ determines also the cavity eigenfrequencies:
\be
\omega_m^2 = \omega_0^2\left(1+\mu m^2\right), \qquad
\omega_0 =\frac{c\pi}{L_x L_y}\sqrt{L_x^2 +L_y^2}.
\label{omegmu}
\ee 
In this case we have
\be
\Psi_0=1, \qquad
\Psi_{n\neq 0} =\left[\left(1 +\mu n^2\right)/2\right]^{-1/2}\cos\left(\pi n L_s/L\right), 
\ee
so  that
\be
 \rho \le 1 +2\sum_{n=1}^{\infty} \left(1 +\mu n^2\right)^{-1} =
\left(\pi/\sqrt{\mu}\right)\coth\left(\pi/\sqrt{\mu}\right),
\label{rhomu}
\ee
where we have used Eq. 6.1.25.4 from \cite{Br}.

\end{document}